\documentclass[aps,pra,twocolumn,superscriptaddress,showpacs]{revtex4}

\usepackage{amsmath}
\usepackage[T1]{fontenc}
\usepackage{newlfont}
\usepackage{amssymb}
\usepackage{amsfonts}
\usepackage{graphicx}
\usepackage{bm}
\usepackage{hyperref}
\newcommand{\ket}[1]{|#1\rangle}
\newcommand{\bra}[1]{\langle#1|}

\renewcommand{\t}[1]{\textrm{#1}}

\newcommand{\sixj}[2]{\left\{ \begin{array}{ccc} #1 \\ #2   \end{array}\right\}}

\begin{document}

\title{Effects of imperfect noise correlations on decoherence-free subsystems:
SU(2) diffusion model}

\author{Rafa{\l} Demkowicz-Dobrza{\'n}ski}
\affiliation{Center for Theoretical Physics of the Polish Academy of Sciences
Aleja Lotnik{\'o}w 32/44, 02-668 Warszawa, Poland}
\author{Piotr Kolenderski}
\affiliation{Instytut Fizyki, Uniwersytet Miko{\l}aja Kopernika, ul.
Grudzi\k{a}dzka 5, 87-100 Toru{\'n}, Poland}

\author{Konrad Banaszek}
\affiliation{Instytut Fizyki, Uniwersytet Miko{\l}aja Kopernika, ul.
Grudzi\k{a}dzka 5, 87-100 Toru{\'n}, Poland}

\date{\today}

\begin{abstract}
We present a model of an $N$-qubit channel where consecutive qubits
experience correlated random rotations. Our model is an extension to
the standard decoherence-free subsystems approach (DFS) which
assumes that all the qubits experience the same disturbance. The
variation of rotations acting on consecutive qubits is modeled as
diffusion on the $\text{SU}(2)$ group. The model may be applied to
spins traveling in a varying magnetic field, or to photons passing
through a fiber whose birefringence fluctuates over the time
separation between photons. We derive an explicit formula describing
the action of the channel on an arbitrary $N$-qubit state. For $N=3$
we investigate the effects of diffusion on both classical and
quantum capacity of the channel. We observe that nonorthogonal
states are necessary to achieve the optimal classical capacity.
Furthermore we find the threshold for the diffusion parameter above
which coherent information of the channel vanishes.
\end{abstract}


\maketitle

\section{Introduction}


A fruitful approach to protect quantum systems from decoherence
introduced by uncontrolled interactions with the environment is to
use symmetries exhibited by those interactions. When elementary
quantum systems in an ensemble are coupled to the environment in an
identical way, it is possible to identify certain collective degrees
of freedom that turn out to be completely decoupled from the
interaction, thus preserving quantum coherence. This approach is the
basic idea behind decoherence-free subspaces and subsystems (DFSs)
\cite{Palma1996,Bartlett2006} (for a review see \cite{Lidar2003}),
which can be implemented in a number of physical scenarios including
quantum dots \cite{Zanardi1998}, atoms in a cavity \cite{Beige2000},
ion traps \cite{Mang2002} or photons transmitted through a
birefringent fiber \cite{Bartlett2003}. The existence of DFSs allows
one to encode or communicate reliably both classical and quantum
information, which has been demonstrated in first proof-of-principle
experiments
\cite{Kwiat2000}.

The purpose of this paper is to go beyond the standard theory of DFSs and to analyze
a scenario when the interactions of elementary systems with the environment are not necessarily identical. Specifically, we will consider a sequence of qubits affected by
random unitary transformations. In the standard DFSs model all unitaries are the same, enabling one to apply the standard decomposition into multiplicity subspaces based
on angular-momentum algebra. In this paper, we will assume that correlations between unitaries affecting consecutive qubits are characterized by a probability distribution describing isotropic diffusion on the SU(2) group. Such a distribution has a number of invariant properties that result in the existence of a single real parameter which defines the strength of correlations. This will enable us to investigate the continuous
transition between the extreme regimes of perfect correlations and independent transformations of consecutive qubits.

The model introduced in this paper
is a certain  channel acting on a sequence of $N$ qubits. We derive its detailed description using angular momentum algebra, and analyze quantitatively the case of $N=3$ qubits. In particular we present
numerical results concerning the classical capacity and present the
optimal ensemble of states attaining it. Surprisingly, for imperfect
correlations the optimal
ensemble contains nonorthogonal states. The fact that the use of
nonorthogonal states may be necessary for the optimal classical
communication has been known \cite{Fuchs1997}, yet it is rather surprising that
this effect appears in a channel constructed in such a natural way.
We compare the optimal capacity with the one that can be achieved
when one is restricted to the use of orthogonal states and
indicate orthogonal states which perform almost optimal.
We calculate and analyze the optimal coherent information, which provides certain
information about the quantum capacity of the channel.

Results presented in this paper can be applied to different physical systems used for quantum communication, e.g.\ spins traveling in fluctuating external magnetic field, and
photons traveling in a fiber with randomly varying birefringence. The model generalizes
the standard DFS theory whenever
the time separation between qubits sent through the channel is not negligible in comparison with the characteristic time of environment fluctuations.
One can easily imagine such a situation in the case of spins traveling relatively slowly in randomly varying magnetic field.
When it comes to photons, however, in most of present experiments the standard DFS approach seems satisfactory. The time separation of photons is typically several
orders of magnitude smaller than the fluctuation time of birefringence, whose variations are caused by  mechanical and thermal
factors and hence by their nature are relatively slow. Nevertheless, our model might be relevant also in this setting,
when large temporal separations between photons need to be introduced, or sequences of photons are long enough to reach the time scale of birefringence fluctuations.

The paper is organized as follows. We
describe the model of imperfect correlations in Sec.~\ref{sec:model}.
In Sec.~\ref{sec:action}
investigate the structure of output
states and give an explicit and efficient formula for calculating the
action of the channel on an arbitrary $N$ qubit state.
In Sec.~\ref{sec:three} we consider the example of three qubit
communication, which is the smallest number of qubits allowing
quantum information to be transmitted. Finally, Sec.~\ref{sec:conclusions}
concludes the paper.

\section{Imperfect correlations}
\label{sec:model}

The standard model for collective depolarization of $N$ qubits
described by a joint density matrix $\rho$ is given by the
following map:
\begin{equation}
\label{eq:twirling}
 \mathcal{T}(\rho)=\int \t{d}U \ U^{\otimes N}
\rho U^{\dagger \otimes N},
\end{equation}
where $U$ is an $SU(2)$ matrix describing the rotation of a Bloch
vector of a single qubit state and $\t{d}U$ is the Haar measure on
the $SU(2)$ group. This operation inflicts a rotation which is
completely random and identical for all qubits. In other words we
may say that noise experienced by a given qubit is perfectly
correlated with the noise experienced by all other qubits. We will
refer to this map as \emph{the twirling map}.

Let us now extend the above model in order to describe the situation
where correlation of noise acting on different qubits is not
perfect. We will assume that the density matrix $\rho$ of the qubits
undergoes a unitary transformation given by a tensor product $U_1
\otimes U_2 \otimes \dots \otimes U_{N}$ averaged with a certain
probability distribution $p(U_1, U_2, \ldots , U_N)$. We will treat
the sequence of $U_i$ as a discrete-time stochastic process with the
Markov property, i.e.\ as a Markov chain \cite{Vankampen1992}.
Consequently the probability distribution for $U_i$ will only depend
on $U_{i-1}$. Additionally, we will impose the stationarity
condition on our process which means that the conditional
probability distribution is described by the same function
irrespectively of the index $i$. Thus the joint probability
distribution is given by a product
\begin{equation}
p(U_1, U_2, \ldots , U_N) = p(U_N|U_{N-1}) \cdots p(U_{3}|U_{2})
p(U_2 | U_1).
\end{equation}
We will consider an isotropic
process, i.e. one that does not distinguish any element of the
$SU(2)$ group, which is equivalent to the physical assumption that
the fluctuations do not favor any particular form of depolarization.
The isotropy
condition requires that the conditional probability
distribution $p(U_i | U_{i-1})=p(V U_i | V U_{i-1})=p(U_i
V | U_{i-1} V)$, for every $V\in \text{SU}(2)$. This implies that
\begin{equation}\label{eq:uni}
    p(U_i | U_{i-1}) = p(U_i U_{i-1}^\dagger)=p(U_{i-1}^\dagger U_i),
\end{equation}
where $p(U)$ is a certain probability distribution defined on the group
SU(2), which fully characterizes our decoherence model.

As the explicit model for the distribution $p(U)$ we will take the
solution to the diffusion equation on the group SU(2). This solution
can be conveniently parameterized with a non-negative dimensionless
time $t$ characterizing the diffusion strength. The explicit
expression for the distribution $p_t(U)$ on the $\text{SU}(2)$ group
at time $t$ reads \cite{Hogan2004}:
\begin{equation}\label{pu}
    p_t(U)=\sum_{j=0}^\infty (2j+1)
    \exp\left(-\frac{1}{2}j(j+1)t\right) \sum_{m=-j}^{j
    }\mathfrak{D}^{j}(U)_m^m,
\end{equation}
where $\mathfrak{D}^j(U)_m^{m^\prime}$ are rotation matrices of the
$SU(2)$ group \cite{Devanthan2002,Tung2003,Brink1968}. The function
$p_t(U)$ is invariant with respect to unitary transformations of its
argument, i.e.\ $p_t(VUV^\dagger)=p_t(U)$ for any $U,V \in
\text{SU}(2)$.
It depends only on the eigenvalues of $U$ which can
be parameterized with a single angle $\xi$ as $e^{i\xi/2}$ and
$e^{-i\xi/2}$. Then the sum over $m$ in Eq.~(\ref{pu}), equal to the
character of the $j$th irreducible representation, can be calculated
explicitly as:
\begin{equation}
\sum_{m=-j}^{j}\mathfrak{D}^{j}(U)_m^m
= \frac{\sin[(j+1/2)\xi]}{\sin(\xi/2)}.
\end{equation}
The family of distributions $p_t(U)$ has the following convolution property:
\begin{equation}
\int \t{d} U \, p_{t_1} (U'U^\dagger)p_{t_2} (U) = p_{t_1+t_2} (U').
\end{equation}
For $t\rightarrow 0$ the function $p_t(U)$ tends to a delta-like
distribution peaked at identity, while for $t\rightarrow\infty$ it becomes
uniform on the group. Therefore the limit $t\rightarrow 0$ corresponds to perfectly
correlated depolarization specified in Eq.~(\ref{eq:twirling}), and
the opposite case of independent depolarization affecting each one of qubits
is recovered when $t\rightarrow\infty$. We will be interested in perturbations in DFSs resulting from imperfect correlations characterized by a finite value of $t$.

\section{Channel action}
\label{sec:action}

We now proceed to write explicitly the action of
the channel on an arbitrary $N$ qubit state $\rho$. The reasoning
below is valid for an arbitrary isotropic function $p(U)$, and the
exact form of $p(U)$ specified in Eq.~(\ref{pu}) is not used until
Eq.~\eqref{eq:evolvefinal}. The output state of the channel reads:
\begin{widetext}
\begin{align*}
    \mathcal{E}(\rho)=\int \t{d}U_1 \int \textrm{d}U_2 \dots \int
    \textrm{d}U_{N} p(U_1,U_2,\dots,U_{N}) U_1 \otimes U_2 \otimes
    \dots \otimes U_{N}\ \rho \
    U_1^\dagger \otimes U_2^\dagger \otimes \dots \otimes U_{N}^\dagger =\\
    = \int \t{d}U_1 \int \textrm{d}U_2 \dots \int \textrm{d}U_{N}
    p(U_2|U_1)p(U_3|U_2)\dots p(U_{N}|U_{N-1}) U_1 \otimes U_2
    \otimes \dots \otimes U_{N}\ \rho \
    U_1^\dagger \otimes U_2^\dagger \otimes \dots \otimes U_{N}^\dagger.\\
\end{align*}
Taking into account Eq.~(\ref{eq:uni}), and introducing new variables
$U_i^\prime=U_i U^\dagger_{i-1}$ we arrive at:
\begin{multline}\label{eq:evolve}
    \mathcal{E}(\rho)=\int \t{d}U_1 \int \textrm{d}U^\prime_2 \dots
    \int \textrm{d}U^\prime_{N} p(U_2^\prime)p(U_3^\prime)\dots
    p(U_{N}^\prime)\times
    \\
    U_1 \otimes (U_2^\prime U_1)  \otimes \dots \otimes
    (U^\prime_{N} U^\prime_{N-1}\dots U_1) \ \rho\ U_1^\dagger
    \otimes (U_1^\dagger U_2^{\prime \dagger}) \otimes \dots \otimes
    (U_1^\dagger U_2^{\prime \dagger} \dots U_{N}^{\prime \dagger})=
    \\
    = \int \t{d}U^\prime_{N} p(U_{N}^\prime)(\openone \otimes \dots
    \otimes U_{N}^\prime) \dots \left(\int \textrm{d}U_1 (U_1
    \otimes \dots \otimes U_1) \ \rho \ (U_1^\dagger \otimes \dots
    \otimes U_1^\dagger) \right) \dots (\openone \otimes \dots
    \otimes U_{N}^{\prime \dagger}).
\end{multline}
\end{widetext}
The action of the channel can thus be written in a compact form:
\begin{equation}\label{eq:x}
    \mathcal{E}(\rho)=\mathcal{I}_{N-1}\left(\mathcal{I}_{N-2}\left(\dots
    \mathcal{I}_{1}(\mathcal{T}(\rho)) \dots\right)\right),
\end{equation}
where:
\begin{multline}\label{eq:difop}
    \mathcal{I}_i(\rho)=\int \textrm{d}U p(U)
    \\
    \times\underbrace{\openone \otimes \openone}_{i} \otimes \underbrace{U
    \otimes \dots \otimes U}_{N-i}\ \rho \ \underbrace{\openone
    \otimes \openone}_{i} \otimes \underbrace{U^\dagger \otimes
    \dots \otimes U^\dagger}_{N-i}.
\end{multline}

It can be easily proven that thanks to the isotropy of the function
$p(U)$ specified in Eq.~\eqref{eq:uni}, operations $\mathcal{I}_i$ commute with
each other, i.e.
$\mathcal{I}_i(\mathcal{I}_k(\rho))=\mathcal{I}_k(\mathcal{I}_i(\rho))$,
and furthermore they commute with $\mathcal{T}$. This implies that
$\mathcal{T}(\mathcal{E}(\rho))=\mathcal{E}(\mathcal{T}(\rho))$,
which together with the fact that
$\mathcal{T}(\mathcal{T}(\rho))=\mathcal{T}(\rho)$ leads us to the
conclusion that $\mathcal{T}(\mathcal{E}(\rho))=\mathcal{E}(\rho)$.
This means that the output state is invariant under the twirling
map. We will say that it has a \emph{twirled structure}.

The action of the twirling map in Eq.~\eqref{eq:x} allows us to
restrict our considerations to input states having a twirled
structure. Let us now describe the states
$\rho_{\t{twirled}}=\mathcal{T}(\rho)$. For this purpose we
use the following standard decomposition of the space of $N$ qubits:
\begin{equation}
\label{eq:decomp}
    \mathcal{H}^{\otimes N} =
    \bigoplus_{j=(N \bmod 2)/2}^{N/2} \underbrace{ \mathcal{H}_j \oplus \ldots \oplus \mathcal{H}_j}_{d_j \t{ times}}
    = \bigoplus_{j=(N \bmod 2)/2}^{N/2} \mathcal{H}_j
    \otimes \mathbb{C}_{d_j},
\end{equation}
where $\mathcal{H}_j$ is a representation subspace in which the
$(2j+1)$-dimensional irreducible representation of $\text{SU}(2)$
acts, and $\mathbb{C}_{d_j}$ is the multiplicity subspace which is not
affected by the action of $U^{\otimes N}$.
The above decomposition allows us to use a basis \mbox{$|j,m,\alpha
\rangle$}, where $j$ is the total angular momentum, $m$ is
projection of angular momentum on the $z$ axis, and $\alpha$
labels representation subspaces that
are equivalent, i.e.\ have identical $j$.

\begin{figure}
  \includegraphics[width=\columnwidth]{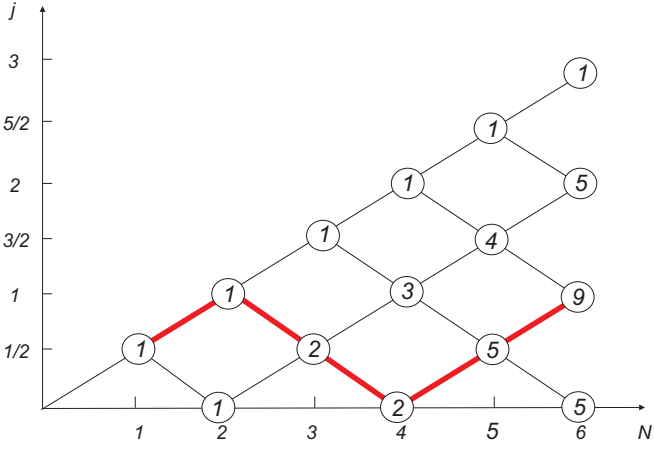}\\
  \caption{Van Vleck's branching diagram describing addition of $N$
spins $1/2$. The $X$ and $Y$ axis corresponds to number of spins $N$
and the total spin $j$ respectively. The numbers in circles
represent the multiplicity of given total angular momentum subspace
(number of paths leading to it). An exemplary path
$\{\frac{1}{2},1,\frac{1}{2},0,\frac{1}{2},1\}$ is marked with bold
(red) line.}\label{fig:vanvleck}
\end{figure}

There are different ways of obtaining a given angular momentum
by adding elementary $1/2$-spins of single qubits. If a given angular momentum is obtained by always adding the $i$-th spin-$1/2$ to the
previously obtained total angular momentum of $i-1$ spins, these different ways correspond to
different paths in the van Vleck diagram \cite{Karwowski2003},
shown in Fig.~\ref{fig:vanvleck}. Therefore $\alpha$ can be regarded as an index labelling different paths leading to equivalent representation subspaces.
There are, however,  alternative ways of specifying
$\alpha$ which depend on the order of adding elementary $1/2$ spins. In the following calculations, it will be helpful to use different conventions for labelling equivalent representations, distinguished with a subscript $\{k\}$. In explicit terms, $\alpha_{\{k\}}$ will label equivalent representation subspaces which are obtained by first adding successively the first $k$ spins $1/2$, from number $1$ to $k$, then adding the last $N-k$ spins in the reverse order, from number $N$ to $k+1$, and finally combining these two groups together. The value of $\alpha_{\{k\}}$ is thus fully specified by a sequence $\{j_1,j_{12},\dots,j_{1\dots k}\}\{j_{k+1\dots N},j_{k+2 \dots N},\dots,j_N\}$, where $j_{i\dots k}$ is the total angular momentum after adding together spins from the number $i$ to $k$. Of course, a single particle spin is equal to $j_i=1/2$ for any $i$. Naturally, all the conventions for numbering equivalent subspaces, corresponding to a different choice of the $k$, are legitimate, yet the ability to switch between alternative conventions will make it possible to derive a compact formula for the action  of the channel.

 Using the decomposition given in Eq.~\eqref{eq:decomp} we can write any
twirled state $\rho_{\t{twirled}}$ in the following way:
\begin{equation}\label{eq:twirledstrucuture}
    \rho_{\t{twirled}}=\mathcal{T}(\rho)= \bigoplus_{j=(N \bmod 2)/2}^{N/2}
    \frac{p_j}{2j+1} \openone_{\mathcal{H}_j} \otimes \rho_j,
\end{equation}
Denoting by $P_j$ a projection onto $\mathcal{H}_j \otimes
\mathbb{C}_{d_j}$ we find the relations
\begin{eqnarray}
    p_j&=&\t{Tr}(P_j \rho),\\
    \rho_j &=& \frac{1}{p_j} \t{Tr}_{\mathcal{H}_j}(P_j \rho P_j).
\end{eqnarray}
Because the action of $\mathcal{I}_i$ preserves the twirled structure,
we may write the output state of the channel in an analogous form:
\begin{equation}
    \mathcal{E}(\rho)=\mathcal{E}(\rho_{\t{twirled}})=
    \bigoplus_{j=(N \bmod 2)/2}^{N/2} \frac{p^\t{out}_j}{2j+1}
    \openone_{\mathcal{H}_j} \otimes \rho^\t{out}_j.
\end{equation}
Therefore, the full description of the action of the channel
amounts to writing $p_j^\t{out}, \rho_j^\t{out}$ as a function of
$p_j, \rho_j$. Equivalently we may characterize the channel by its action
on operators:
\begin{equation}
\begin{split}
\label{Eq:PJalphalpha'}
P_{J}^{\alpha,\alpha^\prime}&=\frac{1}{2J+1}\sum_{M=-J}^{J}
    |J,M,\alpha \rangle \langle J,M, \alpha^{\prime}|=\\
    &=\frac{1}{2J+1} \openone_{\mathcal{H}_{J}} \otimes
    \ket{\alpha}\bra{\alpha^\prime}
\end{split}
\end{equation}
since any twirled state can be written as their linear combination.
As discussed before, we have freedom in choosing the convention for
labels $\alpha$. The simplest expression is obtained if we specify how the
channel acts on $P_J^{\alpha_{\{1\}},\alpha^\prime_{\{1\}}}$, and express
the output in terms of the operators
$P_J^{\alpha_{\{N-1\}},\alpha^\prime_{\{N-1\}}}$. The detailed derivation is
presented in Appendix~\ref{App:derivation}, with the final formula in the form:
\begin{widetext}
\begin{equation}\label{eq:evolvefinal}
\begin{split}
    \mathcal{E}(P_J^{\alpha_{\{1\}},\alpha_{\{1\}}^\prime})&=
    \sum_{\substack{j_{12}^\prime\\j_{12}}} \ \dots\
    \sum_{\substack{j_{1\dots N-1}^\prime\\j_{1 \dots N-1}}}
    \sum_{J_1=|j_1-j_{2\dots N}|}^{j_1+j_{2\dots N}}\
    \sum_{J_2=|j_{12}-j_{3\dots N}|}^{j_{12}+j_{3\dots N}}
\ \dots\
    \sum_{J_{N-1}=|j_{1\dots N-1}-j_N|}^{j_{1\dots N-1}+j_N}\
    P_{J_{N-1}}^{\alpha_{\{N-1\}},\alpha_{\{N-1\}}^\prime}  \times
    \\
    & U(J_1,2)^{j_{12}}_{j_{2\dots N}} U(J_1,2)^{j^\prime_{12}}_{j^\prime_{2\dots N}}\  \dots\  U(J_{N-2},N-1)^{j_{1\dots N-1}}_{j_{N-1,N}} U(J_{N-2},N-1)^{j^\prime_{1 \dots N-1}}_{j^\prime_{N-1,N}} \times
    \\
    & R(t)^{J_1,j_1^\prime,j_{2\dots N}^\prime}_{J, j_1,j_{2\dots N}}
R(t)^{J_2,j_{12}^\prime,j_{3\dots N}^\prime}_{J_1, j_{12},j_{3\dots N}}
    \ \dots \ R(t)^{J_{N-1},j_{1\dots N-1}^\prime,j_N^\prime}_{J_{N-2},j_{1\dots N-1},j_N}
\end{split}
\end{equation}
\end{widetext}
where
according to our convention $\alpha_{\{k\}}$, and
$\alpha_{\{k\}}^\prime$ read:
\begin{equation}
\begin{split}
\alpha_{\{k\}} &= \{j_1,j_{12},\dots,j_{1\dots
k}\}\{j_{k+1,\dots,N},j_{k+2,\dots, N},\dots,j_N\} \\
\alpha_{\{k\}}^\prime &= \{j_1^\prime,j_{12}^\prime,\dots,j_{1\dots
k}^\prime\}\{j_{k+1,\dots,N}^\prime,j_{k+2,\dots,
N}^\prime,\dots,j_N^\prime\},
\end{split}
\end{equation}
the functions $R(t)^{J_k,j_1^\prime,j_2^\prime}_{J, j_1,j_2}$ are
given in terms of Wigner $6j$ coefficients \cite{Devanthan2002} as:
\begin{equation}
\begin{split}
    R(t)^{J_k,j_1^\prime,j_2^\prime}_{J, j_1,j_2}=
    \sum_{j=|j_2-j_2^\prime|}^{j_2+j_2^\prime}(-1)^{J_k-J}
    (2j+1)(2J_k+1) \times
    \\
    \exp\left(-\frac{1}{2}j(j+1)t\right) \sixj{j_1 & J &
    j_2}{j_2^\prime & j & j_1^\prime} \sixj{j_1 & j_1^\prime &
    j}{j_2^\prime & j_2 & J_k},
\end{split}
\end{equation}
and $U(J,k)^{j_{1\dots k}}_{j_{k\dots N}}$ is a shorthand notation
for specific coefficients $U(j_1,j_2,J,j_3;j_{12},j_{23})$ used in
adding three angular momenta \cite{Devanthan2002}, discussed in
Appendix~\ref{App:derivation}):
\begin{multline}
\label{Eq:UJk}
    U(J,k)^{j_{1\dots k}}_{j_{k\dots N}} =
    U(j_{1\dots k-1},1/2,J,j_{k+1\dots N};j_{1\dots
    k},j_{k\dots N}).
\end{multline}
Recall also that all single particle spins $j_i=1/2$, which we kept implicit in Eq.~(\ref{eq:evolvefinal}) only to ease the
understanding of the formula structure.

The formula (\ref{eq:evolvefinal}), despite its lengthy appearance,
enables an
efficient calculation of the channel action even for a large number of
qubits, which would be infeasible by inserting directly the formula
for $p_t(U)$ into Eq.~\eqref{eq:evolve} and performing the
integration.

\section{Three qubits}
\label{sec:three} The case $N=1$ of one qubit is trivial. The
channel acts as completely depolarizing channel, hence its capacity
both classical and quantum is zero. The case $N=2$ of two qubits was
solved in Ref.~\cite{Ball2004}, where the optimal classical capacity
was derived together with the optimal ensemble of states. The
twirling operation in this case produces Werner states
$\rho=(1+c)\openone/4 -c \ket{\psi^-}\bra{\psi^-}$. Imperfect
correlations were introduced in that paper through a
phenomenological shrinking factor $\eta$ multiplying the parameter
$c$. Our general model predicts the same behavior, with the
shrinking factor given explicitly by $\eta = e^{-t}$. Notice also
that quantum capacity of this channel is zero since all multiplicity
subspaces are one dimensional.

We will now investigate the action of the channel for three qubits,
which is the lowest number with non-trivial equivalent
representation subspaces. Adding three spins $1/2$ gives one
subspace with $j=3/2$, which is the fully symmetric subspace, and
two subspaces with $j=1/2$. The two-fold multiplicity of subspaces
corresponding to $j=1/2$ enables one to preserve quantum
superpositions. In the case of perfect correlations $t=0$ the
channel acts as the identity in the multiplicity subspace
\cite{Bartlett2006}, which therefore forms a decoherence-free
subsystem, making it possible to encode one qubit with the fidelity
equal to one. The classical capacity of such a channel is equal to
$\log_2 3$, since we have three completely distinguishable output
states: two orthogonal states of a qubit from the multiplicity
subspace and one state from the fully symmetric subspace. When
correlations between operations acting on consecutive qubits are not
perfect the fidelity of the transmission as well as classical and
quantum capacity will decrease. In this section we will derive
analytical formulas for the fidelity of transmission through the
channel as well as numerical and approximate analytical results for
quantum and classical capacities of the channel.

For three qubits a general twirled state has the form:
\begin{equation}\label{eq:rho3}
    \rho=\frac{p}{2} ( \openone_{\mathcal{H}_{1/2}}\otimes
    \rho_{1/2} ) \oplus \frac{1-p}{4} \openone_{{\mathcal{H}}_{3/2}},
\end{equation}
where $\rho_{1/2}$ is an arbitrary $2 \times 2 $ density matrix, and $0 \le p \le 1$.
Let
us now specify a particular basis in the $j=1/2$ multiplicity
subspace in which the operation ${\cal E}$ will have the simplest form.
Let the state $\ket{0}$ correspond to the equivalent subspace
labeled $\{j=1/2,\alpha=0\}$, obtained by
combining the first two spins $1/2$ together to get the angular momentum
$0$ and then adding the third spin with the total angular
momentum $1/2$, while $\ket{1}$ corresponds to the equivalent
subspace labeled $\{j=1/2,\alpha=1\}$, obtained by combining the
first two spins to get the angular momentum $1$, and then adding the third
spin to produce the total angular momentum $1/2$.
For simplicity
in what follows we will omit the identity operators acting in the
representation subspaces. Instead of
$\openone_{\mathcal{H}_{1/2}}\otimes \rho_{1/2}$ we will simply
write a single qubit state $\rho_{1/2}$, and instead of
$\openone_{\mathcal{H}_{3/2}} $ we will write $\ket{2}\bra{2}$. Our
channel can be thus regarded as effectively a qutrit channel that
preserves no coherence between subspaces $\{\ket{0},\ket{1}\}$ and
$\{\ket{2}\}$. The evolution of the state has the simplest form if
 we introduce the following basis in the qubit subspace:
\begin{equation}
    \ket{e_1}=\frac{|0\rangle + \sqrt{3}|1\rangle}{2}, \quad
    \ket{e_2}=\frac{\sqrt{3}|0\rangle - |1\rangle}{2}.
\end{equation}
In order to describe the action of the channel it is sufficient to
calculate its action on the extreme states whose
convex combinations generate the entire set of twirled states.
The extreme states are $\ket{2}\bra{2}$ and an arbitrary pure state
$\ket{\psi}\bra{\psi}$ in the qubit subspace, which we will parameterize as
$\ket{\psi}=\cos(\theta/2)\ket{e_1}+\sin(\theta/2)e^{i\phi}\ket{e_2}$.
Applying Eq.~(\ref{eq:evolvefinal}) to the three-qubit case yields:
\begin{widetext}
\begin{equation}\label{eq:evolvesym}
    \mathcal{E}(\ket{2}\bra{2})=
    \left(
    \begin{array}{cc}
        \frac{1}{4}\left(1-e^{-2t}\right) & 0 \\
        0 &\frac{1}{12}\left(3-4e^{-t}+e^{-2t}\right)
    \end{array}
    \right)\oplus
    \begin{array}{c}
        \frac{1}{6}\left(3+2e^{-t}+e^{-2t}\right)
    \end{array}
\end{equation}
\begin{equation}
\label{eq:evolvequbit}
\begin{split}
    \mathcal{E}(\ket{\psi}\bra{\psi})=
    \left(
    \begin{array}{cc}
        \frac{1}{4}\left(1+e^{-2t}(1+2\cos\theta)\right)
        &\frac{1}{2}\sin\theta(e^{-t}\cos\phi-ie^{-2t}\sin\phi)
        \\
        \frac{1}{2}\sin\theta(e^{-t}\cos\phi+ie^{-2t}\sin\phi)&
        \frac{1}{12}\left(3+8e^{-t}\sin^2(\theta/2)-e^{-2t}
        (1+2\cos\theta)\right)
    \end{array}
    \right)\oplus
    \\
    \oplus
    \begin{array}{c}
        \frac{1}{6}\left(3-4e^{-t}\sin^2(\theta/2)-e^{-2t}
        (1+2\cos\theta)\right)
    \end{array},
\end{split}
\end{equation}
\end{widetext}
where the output matrix is written in the basis $\ket{e_1}$,
$\ket{e_2}$, $\ket{2}$. This expression will be now used to calculate
quantities characterizing the channel.

\subsection{Fidelity}
Let us first calculate the transmission fidelity of a pure qubit
state encoded in the multiplicity subspace. According to
Eq.~(\ref{eq:evolvequbit}),
with a probability
$\frac{1}{6}[3-4e^{-t}\sin^2(\theta/2)-e^{-2t}(1+2\cos\theta)]$
the input qubit state is removed from the
multiplicity subspace
and transformed into the state $\ket{2}\bra{2}$. We can
consider an effective one qubit channel $\mathcal{E}_{\t{eff}}$ by
replacing the state $\ket{2}\bra{2}$ at the output with the
maximally mixed state in the qubit space, given by $\openone/2$.

Since now we deal with an ordinary one qubit channel, i.e. a one
qubit trace-preserving completely positive map,
we may investigate its action in terms of an affine
map on the Bloch vector. Written in the basis $\ket{e_1}, \ket{e_2}$,
the output Bloch vector
$\mathbf{r}_{\text{out}}$ can be expressed in terms
of the input vector $\mathbf{r}_{\text{in}}$ as:
\begin{equation}\label{eq:channel:reduced}
    \mathbf{r}_{\text{out}}=
    \left(
    \begin{array}{lll}
    e^{-t} & 0 & 0 \\
    0 & e^{-2 t} & 0 \\
    0 & 0 & \frac{2e^{-2 t} + e^{-t}}{3}
    \end{array}
    \right)
    \mathbf{r}_{\text{in}}+
    \left(
    \begin{array}{l}
    0 \\
    0 \\
    \frac{e^{-2 t} - e^{-t}}{3}
    \end{array}
    \right).
\end{equation}
The channel shrinks and translates the initial Bloch sphere. Notice
that the shrinking is not isotropic and is weakest and strongest respectively
in the $x$ and $y$ directions. The translation magnitude initially
increases achieving the maximal value at $t=\ln(2)$ and then vanishes
asymptotically.

\begin{figure}[t]
    \includegraphics[width=\columnwidth]{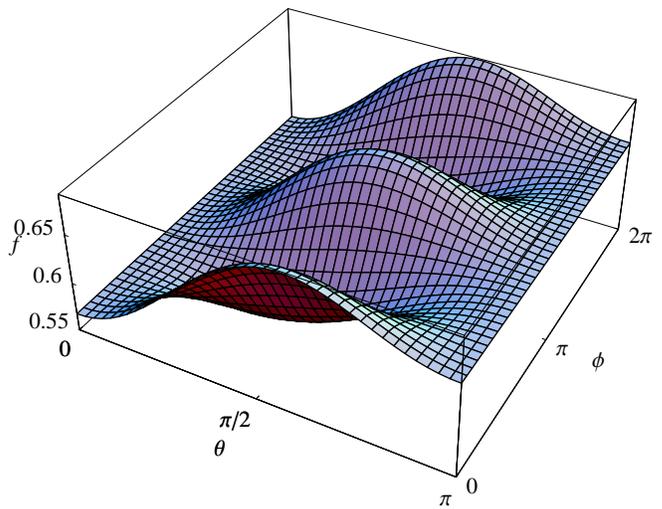}\\
    \caption{The fidelity $f(\theta,\phi,t=1)$
        of  states transmitted through an effective qubit channel
        $\mathcal{E}_\t{eff}$ as a function of the Bloch parameters $\theta$
        and $\phi$ of the input state. }
    \label{fig:fidelity}
\end{figure}

The explicit expression for the fidelity of the output state
is given by:
\begin{multline}\label{fidelity}
    f(\theta,\phi,t)=\bra{\psi}\
    \mathcal{E}_\t{eff}(\ket{\psi}\bra{\psi})\  \ket{\psi}
    =\frac{1}{2}(1+\mathbf{r}_\mathbf{\text{in}} \cdot
    \mathbf{r}_{\text{out}})= \\
 = \frac{1}{24} [ 12 + 5e^{-t} + 7
    e^{-2t}+
        \\
    +(e^{-2t} - e^{-t}) (4 \cos \theta -6 \cos 2 \phi
     \sin ^2\theta + \cos 2 \theta ) ].
\end{multline}
and is depicted in Fig.~\ref{fig:fidelity}. It is seen that the
fidelity is optimized along meridians $\phi=0,\pi$, which is a
consequence of the fact that shrinking is weakest in the $x$
direction.  The highest transmission fidelity is achieved for states
$\sqrt{5/8}\ket{e_1} \pm \sqrt{3/8}\ket{e_2}$, independently of the
actual value of the diffusion time $t>0$.

We now compute the average fidelity of states lying on a great
circle parameterized by the unit normal vector given in spherical
coordinates $(\theta_{c},\phi_{c})$. This quantity is relevant when
one needs to transmit a relative phase of an equally weighted superposition \cite{Huelga2002}. The
average great circle fidelity reads:
\begin{multline}
    f_{c}(\theta_{c},\phi_{c})= \\
    \frac{1}{24} \left[6\left(e^{-2t}+ e^{-t} +2\right) + \left(e^{-2t} - e^{-t}\right)(1+3\cos 2 \phi_{c})\sin^2\theta_{c}
    \right]
\end{multline}
Thus it is a straightforward observation that the great circle
parameterized by $(\theta_c=\pi/2,\phi_c=\pi/2)$ (the meridian in the
$xz$ plane) is optimal and yields $f_c=(3+ 2e^{-t}+e^{-2t})/6$.

The average fidelity integrated over the entire Bloch sphere of
input states with the uniform distribution is given by:
\begin{equation}\label{eq:fidelity:average}
    \langle f(t)\rangle=\frac{1}{18} \left(9 + 4 e^{-t}+ 5 e^{-2
    t}\right)
\end{equation}
and it decreases monotonically with increasing diffusion
strength. In the limit of strong diffusion $t\gg 1$ the the channel
is completely depolarizing.

\subsection{Coherent information}
We will now optimize the coherent information of the
channel $\mathcal{E}$ in order to estimate its capacity for
transmitting quantum information. Let us note that we now need to consider
the full
qutrit channel $\mathcal{E}$ acting according to Eqs.~(\ref{eq:evolvesym})
and (\ref{eq:evolvequbit}) rather than the
effective one qubit channel $\mathcal{E}_\t{eff}$, as considering only the latter
would lower the achievable quantum
capacity.

Coherent information
 is defined as follows \cite{Nielsen2000}:
\begin{equation}\label{eq:cohinf}
I_C=\sup_{\rho}\left(S(\mathcal{E}(\rho))-S_{\t{env}}(\mathcal{E},\rho)\right),
\end{equation}
where $S(\rho)=-\text{Tr}\left(\rho\log_2\rho\right)$ is the von
Neumann entropy and $S_{\t{env}}$ is the entropy exchange
\cite{Nielsen2000}. The analytical optimization seems hard due to
the complicated form of the Krauss operators \cite{Nielsen2000} our
channel has. Nevertheless, one can easily notice that the optimal
state $\rho$ in Eq.~\eqref{eq:cohinf} will be supported on
$\{\ket{e_1},\ket{e_2}\}$ subspace, since the symmetric subspace
$\ket{2}$ can be regarded as a purely classical degree of freedom
and therefore cannot contribute to quantum capacity. Numerical
optimization shows that the optimal state has the form:
\begin{equation}\label{eq:ci:input}
    \rho(\epsilon)=\epsilon\ket{e_1}\bra{e_1}+ (1-\epsilon)
    \ket{e_2}\bra{e_2}.
\end{equation}
We have optimized analytically the coherent information in the limit
of weak diffusion (small $t$). In the case of no diffusion the state
$\rho(\epsilon=1/2)$ maximizes the quantum capacity: $I_C=1$. We
have expanded $I_C$ in a power series in $\epsilon$ around
$\epsilon=1/2$ up to the second order
 and calculated the maximum. In the lowest order in $t$ (which
includes also terms $t\log t$, reflecting the fact that the
derivative of $I_C(t)$ diverges in $t=0$) the approximate input state
parameter $\epsilon$ and the quantum capacity read:
\begin{eqnarray}
\label{eq:icapp}
    I_C &\approx& 1- \frac{t}{3}\left(8 -\log_2 3 + \frac{2}{\ln
    2}- 2 \log_2 t\right)
    \\
\label{eq:icepsilon}
    \epsilon &\approx& \frac{1}{2} + \frac{t}{6}  \left(1 -
    \frac{1}{ \log_2 3} \right)
\end{eqnarray}
\begin{figure}[t]
    \includegraphics[width=\columnwidth]{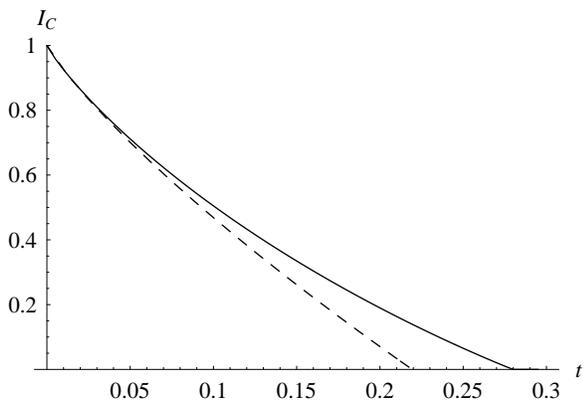}\\
    \caption{Coherent information $I_C$ as a function of diffusion time $t$.
    The numerical solution is marked with the solid line, the analytical
    approximation given by Eq.~\eqref{eq:icapp} is marked with the dashed line.
    } \label{fig:ci}
\end{figure}
Numerical results depicted in Fig.~\ref{fig:ci} indicate that coherent

information drops to strictly zero for diffusion time $t>0.275$,
which strongly suggest that in this regime no quantum communication
is possible.
 Since our channel is not
degradable \cite{Devetak2005} it could happen that the quantum
capacity is not zero even though the coherent information vanishes,
see e.g.\ Ref~\cite{Barnum1998}.
\begin{figure}[t]
    \includegraphics[width=\columnwidth]{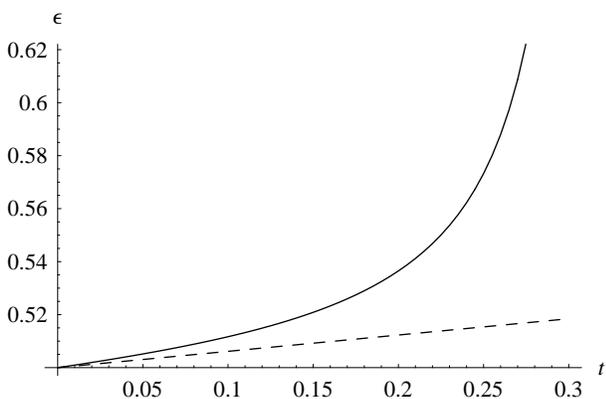}\\
    \caption{The parameter $\epsilon$ of the input state maximizing
    the coherent information $I_C$. The numerical solution is marked
    with the solid line, the analytical approximation given in Eq.
    \eqref{eq:icepsilon} is marked with the dashed line. }
    \label{fig:epsweigh}
\end{figure}

\subsection{Classical capacity}
The classical capacity of a quantum channel can be calculated using
the Holevo-Schumacher-Westmoreland \cite{Schumacher1997,
Nielsen2000} formula:
\begin{equation}\label{eq:capacity}
    C=\sup_{\{p_i,\rho_i\}} \left[ S\left(\mathcal{E}\left(\sum_i
    p_i \rho_i\right)\right)-\sum_i p_i
    S(\mathcal{E}\left(\rho_i)\right)\right],
\end{equation}
where the supremum is taken over all ensembles $\{p_i,\rho_i\}$. Let
$\tilde C =S\left(\mathcal{E}\left(\sum_i p_i \rho_i \right) \right)
- \sum_i p_i S(\mathcal{E}\left(\rho_i)\right)$ denote the
expression that is optimized in Eq.~\eqref{eq:capacity}. In order to
achieve the supremum it is enough to use $d^2$ pure states, where
$d$ is the dimension of the input Hilbert space
\cite{Verstraete2002}. In our case the maximum number of pure stares
needed to achieve the supremum is five: four in the qubit subspace
and additionally the symmetric state $\ket{2}$. We implemented
numerical optimization of the classical capacity, which for all
diffusion times $t$ yielded three-element optimal ensembles
containing only two states from the qubit subspace. Furthermore, the
sub-ensemble in the qubit subspace was composed of states given by:
\begin{equation}
\label{eq:optstates}
\begin{split}
    \ket{\psi_1} &= \cos(\theta/2)\ket{e_1}+\sin(\theta/2)\ket{e_2} \\
    \ket{\psi_2} &= \cos(\theta/2)\ket{e_1}-\sin(\theta/2)\ket{e_2}
    \end{split}
\end{equation}
and equal weights, which we will denote by $q$.
The parameters $q$ and $\theta$ are functions of $t$.

\begin{figure}[t]
    \includegraphics[width=\columnwidth]{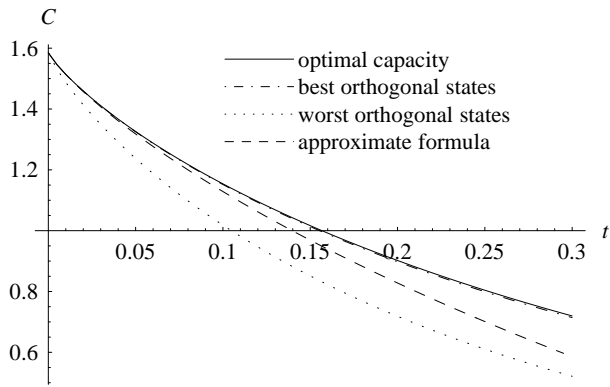}\\
    \caption{The classical capacity of the quantum channel $C$ as a
    function of diffusion time $t$. The solid line depicts numerical
    results, the dashed line depicts the approximate analytical solution
    from Eq.~\eqref{eq:holevo:analit}, while dashed-dotted and dotted
    lines represent capacities obtained using respectively the best and the worst
    pairs of orthogonal states. Notice that using proper orthogonal
    states allows one almost to achieve the optimal capacity.}
    \label{fig:capacity}
\end{figure}

For $t>0$ the optimal states $\ket{\psi_1}$ and $\ket{\psi_2}$
turn out to be non-orthogonal. In order to judge
how significant this nonorthogonality is for the channel transmission,
let us compare the optimal capacity with the capacity
attainable when composing the input ensemble from a pair of
orthogonal states in the qubit
subspace and the symmetric state $\ket{2}$. The optimal orthogonal states for transmission in the qubit subspace
are $1/\sqrt{2}(\ket{e_1}\pm \ket{e_2})$. As seen in Fig.~\ref{fig:capacity},
the capacity obtained using these states is very close to the absolute optimum.
However, a different choice of orthogonal states can significantly
deteriorate the capacity, with the worst case corresponding to the states
$1/\sqrt{2}(\ket{e_1}\pm i\ket{e_2})$. The best and the worst performance of these pairs
can be explained intuitively by the anisotropy of the
effective qubit channel ${\cal E}_\t{eff}$:
the optimal orthogonal states lay on the
$x$ axis characterized by the weakest shrinking,
while the worst ones on the $y$ axis where the shrinking is strongest.
This argument is of course not rigorous, as we are dealing here
with the full qutrit channel rather than the effective one ${\cal E}_\t{eff}$
in the qubit subspace.
While in the limit $t=0$ of no diffusion  any pair
of orthogonal states performs equally well, this invariance is broken for $t>0$ owing
to the asymmetric character of the channel. However, the particular choice of the
orthogonal states given by $1/\sqrt{2}(\ket{e_1}\pm
\ket{e_2})$ performs close to optimum for any diffusion strength.

In order to obtain an approximate analytical solution we expand $\tilde
C= \tilde C(\theta,q,t)$ in a power series up to the second order in
$(\theta,q)$ around $(\pi/2,1/3)$ and optimize analytically. This yields an
approximate solution for a weak diffusion, given by:
\begin{eqnarray}
    q &\approx&  \frac{1}{3} + \frac{t}{108}  \left(5 + \frac{7}{4 \log_2
    3} + \ln t\right)\label{eq:holevo:optimal:q}
    \\
    \theta &\approx& \frac{\pi }{2}- \frac{t}{12}  \left(1- \log_2 3 + 2
    \ln t \right)\label{eq:holevo:optimal:theta}
\end{eqnarray}
\begin{figure}[t]
    \includegraphics[width=\columnwidth]{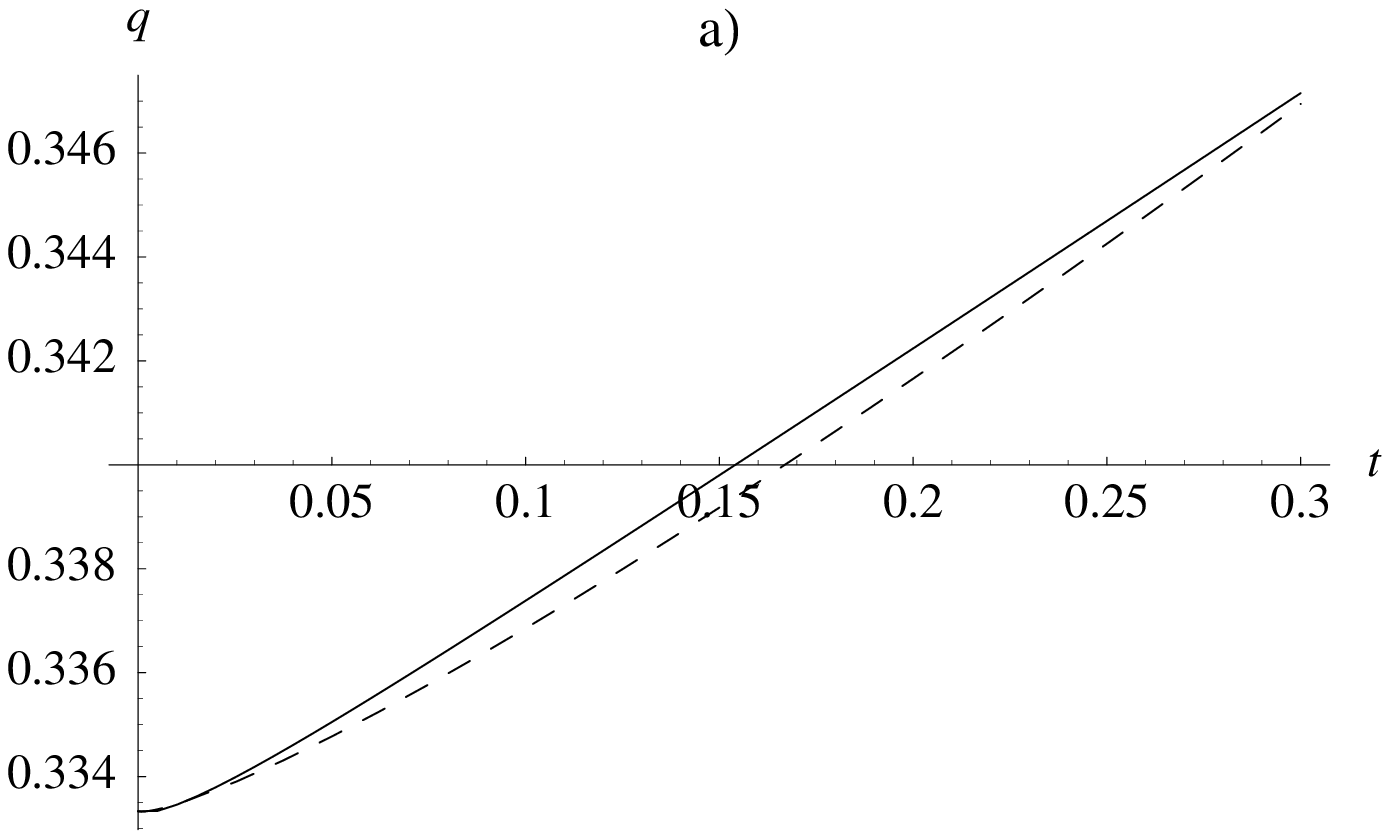}\\
    \includegraphics[width=\columnwidth]{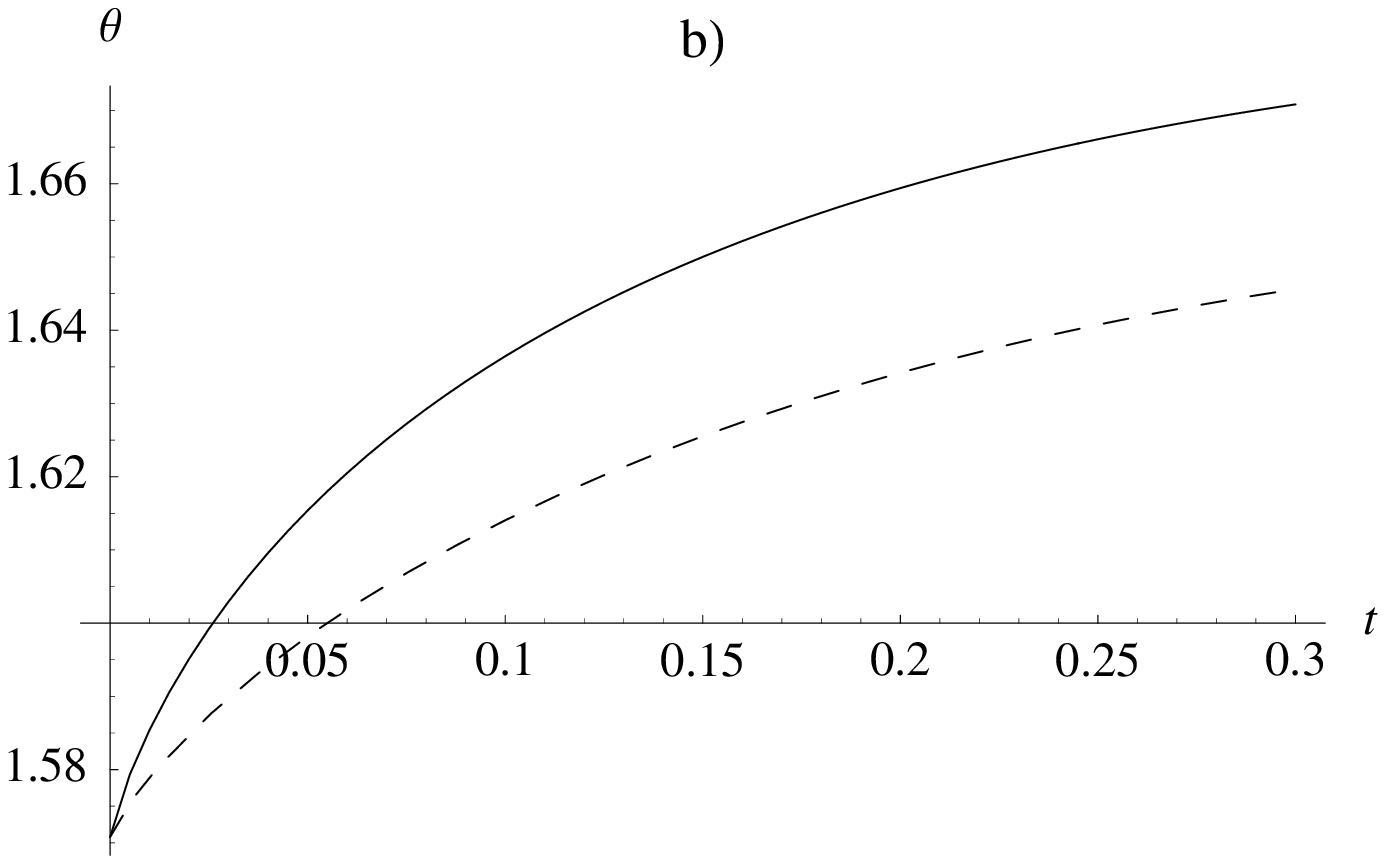}\\
    \caption{The optimal weight $q$ and the optimal Bloch parameter $\theta$, given in Eq.~\eqref{eq:optstates}, of input states belonging to the
    $j=1/2$ multiplicity subspace maximizing classical information  transmission.
    The solid lines depict numerical results, dashed lines
    represent approximate analytical solutions, given in Eqs.~\eqref{eq:holevo:optimal:q} and \eqref{eq:holevo:optimal:theta}.
    Notice that for $t>0$ the states are nonorthogonal.}
    \label{fig:qandtheta}
\end{figure}

A comparison between numerical and approximate analytical solutions is
shown in Fig.~\ref{fig:qandtheta}. The graphs show a
good agreement of analytical solution with numerical results.
Substituting Eqs.~\eqref{eq:holevo:optimal:q} and
\eqref{eq:holevo:optimal:theta} into $\tilde C(\theta,q,t)$ and
retaining leading terms yields an approximate formula for the
capacity:
\begin{multline}\label{eq:holevo:analit}
        C \approx \log_2 3 +\frac{t}{9} \left( 1 - \frac{14 +11\ln 3
        }{2 \ln 2}+ 7 \log_2 t\right),
\end{multline}
which is compared with numerical results in Fig.~\ref{fig:capacity}.

\section{Conclusions and discussion}
\label{sec:conclusions}

In this paper we have introduced a model an $N$ qubit channel with
imperfectly correlated noise, i.e. rotations inflicting consecutive
qubits are subjected to the process of diffusion. We have given an explicit formula for the action of the channel on an arbitrary $N$ qubit state and
for $N=3$ we calculated the optimal classical and quantum capacities and the states which are optimal for communication. Interestingly, we have observed that classical capacity is maximized when using nonorthogonal states. Additionally, we analyzed the robustness of different orthogonal states, which in the case of no diffusion (perfect noise
correlation) are optimal for classical communication. We have indicated the most robust orthogonal states which perform almost optimal for all diffusion times.
We have also found a threshold for the diffusion time above which the coherent information is zero and hence most probably no quantum information can be transmitted.

The model is a very natural extension of the standard DFS theory,
and it can be applied to any physical situation when the action of
the environment on the qubits can be described by a stationary
Markov chain of $\text{SU}(2)$ matrices, and where the transition
probability is described by an isotropic diffusion process on the
$\text{SU}(2)$ group. This is equivalent to the assumption that
consecutive matrices are the result of an isotropic random walk on
the $\text{SU}(2)$ group (see \cite{Laurent2002} for a review on
random walks on groups). The assumption can be justified for systems
we discussed in the introduction: spins traveling in the presence of
randomly varying magnetic field and photons transmitted through a
fiber with birefringence fluctuations. In both cases consecutive
qubits (spins or photons) experience varying rotations in the
channel, caused by fluctuations of the magnetic field or by
birefringence. These multiple small random contributions lead to the
random walk of the effective matrix describing the action of the
channel on consecutive qubits. If these fluctuations are isotropic
then obviously the random walk is isotropic and it leads to the
model we have presented. Furthermore, even if fluctuations are
anisotropic the effective random walk will be isotropic thanks to
the assumption that the channel is long enough to completely
depolarize every single qubit transmitted. This remark is relevant
for fibers where the birefringence fluctuations are typically
modeled as anisotropic \cite{Wai1995}, with the principal axis
corresponding to a linear polarization. However, if such a
fluctuation occurs at some intermediate point of a sufficiently long
fiber, random polarization rotations introduced by the preceding and
the following sections of the fiber should make such fluctuations
isotropic.

\begin{acknowledgments}
P.K. acknowledges insightful discussions with J. Karwowski.
This work has been supported by the Polish
Ministry of Science and Higher Education under grant No
1~P03B~129~30, the European Commission under the Integrated Project Qubit Applications (QAP) funded by the IST directorate as Contract Number 015848, and AFOSR under grant
number FA8655-06-1-3062.

\end{acknowledgments}

\appendix
\section{Derivation of the action of the channel $\mathcal{E}$ on
operators $P_{J}^{\alpha,\alpha^\prime}$} \label{App:derivation}

According to Eq.~\eqref{eq:x} the action of the channel can be expressed
as a composition of operations $\mathcal{I}_i$ defined in Eq.~\eqref{eq:difop}. Let us calculate the action
of $\mathcal{I}_i$ on operators $P_{J}^{\alpha,\alpha^\prime}$ introduced
in Eq.~(\ref{Eq:PJalphalpha'}).
Since $\mathcal{I}_i$ acts non trivially only on the last $N-i$
qubits it will be convenient to number equivalent representation
subspaces using the convention described in
Sec.~\ref{sec:action}, and write explicitly $\alpha_{\{i\}} =
\{j_1,j_{12},\dots,j_{1\dots i}\}\{j_{i+1,\dots,N},j_{i+2,\dots,
N},\dots,j_N\}$. The first step is to calculate the action of
$\mathcal{I}_i$ on $P_J^{\alpha_{\{i\}},\alpha^\prime_{\{i\}}}$:
\begin{multline}
\label{eq:ipi}
\mathcal{I}_i(P_J^{\alpha_{\{i\}},\alpha^\prime_{\{i\}}})=
\int \textrm{d}U p_t(U)\times
    \\
    \underbrace{\openone \otimes \openone}_{i} \otimes \underbrace{U
    \otimes \dots \otimes U}_{N-i}\
    P_J^{\alpha_{\{i\}},\alpha^\prime_{\{i\}}}\ \underbrace{\openone
    \otimes \openone}_{i} \otimes \underbrace{U^\dagger \otimes
    \dots \otimes U^\dagger}_{N-i}.
\end{multline}
Using $\alpha_{\{i\}}$ for labelling equivalent
representations allows
us to decompose the state $\ket{J,M,\alpha_{\{i\}}}$ using
Clebsch-Gordan coefficients, denoted here with square brackets,
according to:
\begin{equation}\label{eq:Clebsch}
\begin{split}
    |J,M,\alpha_{\{i\}} \rangle &= \sum_{m_{1\dots i}=-j_{1\dots i}}^{j_{i+1\dots N}}
    \left[\begin{array}{ccc} j_{1\dots i} & j_{i+1\dots N} & J \\
    m_{1\dots i} & m_{i+1\dots N}& M
    \end{array} \right] \times \\
    &\underbrace{|j_{1\dots i},m_{1 \dots i},\alpha_1 \rangle
    }_{i \t{ first qubits}} \otimes
    \underbrace{|j_{i+1\dots N},m_{i+1 \dots N},\alpha_2\rangle
    }_{N-i \t{ last qubits}}.
\end{split}
\end{equation}
The total angular momenta of the first $i$ and the last $N-i$ qubits,
given respectively by $j_{1\dots i}$ and $j_{i+1\dots N}$ are
uniquely determined by $\alpha_{\{i\}}$. Similarly, the labelling of
equivalent subspaces $\alpha_1$, $\alpha_2$ within each block of
qubits is determined by $\alpha_{\{i\}}$ via the following
relations: $\alpha_1=\{j_1,j_{12},\dots,j_{1\dots i-1}\}\{j_i\}$,
$\alpha_2=\{j_{i+1}\}\{j_{i+2\dots N},\dots,j_N \}$ (remember that
$j_i=1/2$ for any $i$). An analogous
decomposition can be applied to $|J,M,\alpha^\prime_{\{i\}} \rangle$.

The next step is to write the operator
$P_J^{\alpha_{\{i\}},\alpha^\prime_{\{i\}}}$ using the decomposition
given in Eq.~(\ref{eq:Clebsch}). The operation $\mathcal{I}_i$ acts
trivially on the first $i$ qubits, while the action of $U^{\otimes
(N-i)}$ on the last $N-i$ qubits can be written with the help of
Wigner rotation matrix $\mathfrak{D}^j(U)^{m^\prime}_m$. By
substituting the explicit form of $p_t(U)$ given in Eq.~\eqref{pu}
into Eq.~\eqref{eq:ipi} and using properties of Clebsch-Gordan
coefficients, Wigner rotation matrices, and $6j$ Wigner symbols (see
chapters 3,5,7 in Ref.~\cite{Devanthan2002}) we arrive after some
lengthy calculations at a compact formula:
\begin{equation}
    \label{eq:diffstep}
\mathcal{I}_i(P_{J}^{\alpha_{\{i\}},\alpha^\prime_{\{i\}}}) =
\sum_{J_i=|j_{1\dots i}-j_{i+1\dots N}|}^{j_{1 \dots i}+j_{i+1 \dots
N}} R(t)^{J_i,j_{1\dots i}^\prime,j_{i+1\dots N}^\prime}_{J,
j_{1\dots i},j_{i+1 \dots N}}
P_{J_i}^{\alpha_{\{i\}},\alpha^\prime_{\{i\}}},
\end{equation} where:
\begin{equation}
\begin{split}
    R(t)^{J_i,j_1^\prime,j_2^\prime}_{J, j_1,j_2}=
    \sum_{j=|j_2-j_2^\prime|}^{j_2+j_2^\prime}(-1)^{J_i-J}
    (2j+1)(2J_i+1) \times
    \\
    \exp\left(-\frac{1}{2}j(j+1)t\right) \sixj{j_1 & J &
    j_2}{j_2^\prime & j & j_1^\prime} \sixj{j_1 & j_1^\prime &
    j}{j_2^\prime & j_2 & J_i}
\end{split}
\end{equation}
and the curly brackets denote Wigner $6j$ symbols.
Notice that the expression derived in Eq.~(\ref{eq:diffstep})
can be used only if the index $i$
in the operation $\mathcal{I}_i$ is identical with the index $i$
specifying the
convention for numbering equivalent subspaces $\alpha_{\{i\}}$.
If we want to apply Eq.~(\ref{eq:diffstep}) to calculate the full
action of the channel, given by
\begin{equation}
    \mathcal{E}(P_J^{\alpha,\alpha^\prime})=
    \mathcal{I}_{N-1}(\dots \mathcal{I}_{2}(
    \mathcal{I}_{1}(P_J^{\alpha,\alpha^\prime}))
    \dots)
\end{equation}
it is convenient to start from $P_J^{\alpha_{\{1\}},\alpha^\prime_{\{1\}}}$
for the representation of the input state,
and then to adjust the convention after each step. To carry on this procedure we need
to be able to express operators $P_J^{\alpha_{\{i-1\}},\alpha^\prime_{\{i-1\}}}$ in terms
of $P_J^{\alpha_{\{i\}},\alpha^\prime_{\{i\}}}$.

The necessary expression can be derived using the standard formalism for adding
three angular momenta. Consider three spins $j_1$, $j_2$, and
$j_3$. One can write a basis using the total angular momentum of all
spins in two different ways depending on the order in which spins
were added together. A ket $|j_1,(j_2 j_3) j_{23},J,M \rangle$
corresponds to a state with the total angular momentum $J$ and
projection on the $z$ axis $M$, when spins $j_2$, $j_3$ were first
coupled together yielding the angular momentum $j_{23}$ and finally
the spin $j_1$ was added resulting in the total angular momentum
$J$. Analogously $|(j_1 j_2)j_{12}, j_3 ,J,M \rangle$ corresponds to
the situation when first spins $j_1$ and $j_2$ are added and
subsequently the spin $j_3$ joins them. A unitary operation
$U(j_1,j_2,J,j_3;j_{12},j_{23})$ that transforms between these bases:
\begin{multline}
    |j_1 ,(j_2 j_3) j_{23},J,M \rangle =
    \sum_{j_{12}=|j_1-j_2|}^{j_1+j_2} U(j_1,j_2,J,j_3;j_{12},j_{23}) \\
    \times
    |(j_1 j_2) j_{12},j_3,J,M \rangle
\end{multline}
can be expressed using $6j$ Wigner symbols as:
\begin{multline}\label{eq:sixj}
    U(j_1,j_2,J,j_3;j_{12},j_{23})= \sqrt{(2j_{12}+1)(2j_{23}+1)} \\
    \times
    (-1)^{-(j_1+j_2+J+j_3)}
    \sixj{j_1& j_2 & j_{12}}{j_3 &J& j_{23}}
\end{multline}
Specializing these general formulas to our calculation (i.e. replacing $j_1$ with $j_{1\dots i-1}$, $j_2$ with $1/2$ and $j_3$ with $j_{i+1,\dots N}$) we can write:
\begin{multline}
\label{eq:reexpress}
    P_{J}^{\alpha_{\{i-1\}},\alpha_{\{i-1\}}^\prime}=
    \sum_{\substack{j_{1\dots i}^\prime\\j_{1\dots i}}} U(J,i)^{j_{1\dots
    i}}_{j_{i\dots N}}
    P_{J}^{\alpha_{\{i\}},\alpha_{\{i\}}^\prime}
    U(J,i)^{j^\prime_{1\dots i}}_{j^\prime_{i \dots N}}.
\end{multline}
where the coefficients $U(J,k)^{j_{1\dots k}}_{j_{k\dots N}}$
are defined in Eq.~(\ref{Eq:UJk}). For completeness let us also write an inverse relation allowing for the lowering of the index $i$:
\begin{multline}
\label{eq:reexpressinv}
    P_{J}^{\alpha_{\{i\}},\alpha_{\{i\}}^\prime}=
    \sum_{\substack{j_{i\dots N}^\prime\\j_{i \dots N}}} U(J,i)^{j_{1\dots
    i}}_{j_{i\dots N}}
    P_{J}^{\alpha_{\{i-1\}},\alpha_{\{i-1\}}^\prime}
    U(J,i)^{j^\prime_{1\dots i}}_{j^\prime_{i \dots N}}.
\end{multline}

Equipped with the above formulas we can now calculate the
action of the complete channel. For the input state expressed as a combination of
operators $P_{J}^{\alpha_{\{1\}},\alpha_{\{1\}}^\prime}$,
the action of the operation ${\cal I}_{1}$ is, according to Eq.~(\ref{eq:diffstep})
given by:
\begin{multline}
    \mathcal{I}_{1}(P_{J}^{\alpha_{\{1\}},\alpha_{\{1\}}^\prime})=
    \sum_{J_1=|j_1-j_{2\dots N}|}^{j_1+j_{2\dots N}}
    R(t)^{J_1,j^\prime_1,j_{2\dots N}^\prime}_{J, j_1,j_{2\dots N}}
    P_{J_1}^{\alpha_{\{1\}},\alpha_{\{1\}}^\prime}.
\end{multline}
In order to calculate the action of $\mathcal{I}_{2}$ we need to
represent $P_{J}^{\alpha_{\{1\}},\alpha_{\{1\}}^\prime}$ in terms of
operators $P_{J}^{\alpha_{\{2\}},\alpha_{\{2\}}^\prime}$.
Using Eq.~(\ref{eq:reexpress}) yields:
\begin{multline}
    P_{J}^{\alpha_{\{1\}},\alpha_{\{1\}}^\prime}=
    \sum_{\substack{j_{12}^\prime\\j_{12}}} U(J,2)^{j_{12}}_{j_{2\dots N}}
    P_{J}^{\alpha_{\{2\}},\alpha_{\{2\}}^\prime}
    U(J,2)^{j^\prime_{12}}_{j^\prime_{2\dots N}}.
\end{multline}
We may now apply the operation $\mathcal{I}_{2}$, whose action
on the operators
$P_J^{\alpha_{\{2\}},\alpha_{\{2\}}^\prime}$ is again given
by Eq.~(\ref{eq:diffstep}). The combined action of
$\mathcal{I}_{2}$ and $\mathcal{I}_{1}$ thus reads:
\begin{widetext}
\begin{equation}
\begin{split}
    \mathcal{I}_{2}(\mathcal{I}_{1}(P_J^{\alpha_{\{1\}},\alpha_{\{1\}}^\prime}))&=
    \sum_{J_1=|j_1-j_{2\dots N}|}^{j_1 + j_{2\dots N}}
    R(t)^{J_1,j_1^\prime,j_{2\dots N}^\prime}_{J,j_1, j_{2\dots N}}
    \sum_{\substack{j_{12}^\prime \\ j_{12}}}
    U(J_1,2)^{j_{12}}_{j_{2\dots N}} U(J_1,2)^{j^\prime_{12}}_{j^\prime_{2\dots N}}
      \sum_{J_2=|j_{12}-j_{3 \dots N}|}^{j_{12}+j_{3 \dots N}} R(t)^{J_2,j_{12}^\prime,j_{3\dots N}^\prime}_{J_1, j_{12},j_{3 \dots N}} \
    P_{J_2}^{\alpha_{\{2\}},\alpha_{\{2\}}^\prime}
\end{split}
\end{equation}
\end{widetext}
Iterating this procedure yields the explicit formula for the action
of the channel ${\cal E}$ given in Eq.~\eqref{eq:evolvefinal}.
Notice in this formula the output state is expressed in terms of
operators $P_J^{\alpha_{\{N-1\}},\alpha_{\{N-1\}}^\prime}$. If
desired this expression can be converted back to the representation
of operators $P_J^{\alpha_{\{1\}},\alpha_{\{1\}}^\prime}$ by
applying repeatedly Eq.~(\ref{eq:reexpressinv}).

\end{document}